\documentclass[]{IEEEphot}  
\jmonth{November}
\pubyear{2019}
\usepackage{graphicx}

\begin{document}

\title{Enhancing Room Security and Automating Class Attendance Using ID Cards}

\author{Shravan Bhat – 171EE240, Nithin R – 171EC131, Pranav S - 171EC135\\ 
}


\maketitle

\begin{receivedinfo}%

\end{receivedinfo}

\begin{abstract}
With the rapid advancements in technology, automation has emerged as the future of human endeavors. From simple tasks like attendance management to complex security systems, automation has the potential to revolutionize various aspects of our lives. This research paper explores the implementation of a method aimed at enhancing room security in hostels and automating class attendance using ID cards. In this study, we propose a system that utilizes the unique identity information stored in ID cards for various security and check-in tasks. By integrating RFID (Radio-Frequency Identification) reader technology, GSM modules, Node MCU, and Arduino, we create a comprehensive solution. The RFID reader scans the ID card, extracting the relevant information and verifying the user's identity. The data is then transmitted via the GSM module to a central database, ensuring real-time monitoring and security measures. Moreover, the system also enables the automation of class attendance. By utilizing the same ID cards, students can simply tap their cards on a reader placed in the classroom. This information is recorded automatically, eliminating the need for manual attendance taking and reducing errors and time consumption. This research project highlights the practical implementation of ID card technology to enhance room security in hostels and automate class attendance processes. By leveraging the power of automation, we aim to streamline administrative tasks, improve security measures, and optimize efficiency in educational institutions and other relevant settings.
\end{abstract}


\section{Introduction}

\large Security and privacy is a basic need for any human being. India's population has been increasing exponentially since 19\textsuperscript{th} century. Hence student intake for colleges has been increasing every year. Automation would help in trivial tasks like taking attendance, or making payments in a locality. Privacy and security is also an issue in many colleges. Adding layers of security to rooms and safe box would prevent petty theft from happening.\newline\newline
Main motivation of this project is to establish a attendance system within our college campus, a cash-less payment system and also to implement safer and key-less room locking systems in our university.

\section{ {\textbf{Literature Survey}}}
\subsection{Survey of State of Art}
Smart card based door lock systems which are expensive and less secure are currently available like the NFC (Near field communication) cards which are used in the hotel rooms. Using these might be very expensive as it requires complex hardware.\newline\newline
Automated attendance are available, which uses finger print as the ID, But implementing that on a large scale like college is difficulty and would come out to be rather expensive.

\subsection{{Features}}
\begin{itemize}
\item RFID card and RFID reader is included in the door lock system. The door unlocks only when the authorized card is scanned and corresponding pin in entered using the keypad provided.
\item The locking and unlocking of the door latch is implemented using servo motors, stepped motors and gears.
\item When a card is scanned an alert SMS is sent to the registered phone number and also an alert notification is generated in the app. When an authorized card is scanned without the user’s consent, the user can shut down the system by sending a message from his phone.
\item The same RFID card can be used in classrooms as a check in attendance system \\ \\
\end{itemize}

\section{ {\textbf{Details of implementation}}}

\subsection{Components Used}
\begin{itemize}
\item Sim900 GSM module
\item Arduino Uno
\item MFRC522 RFID reader and RFID cards
\item Servo motors, stepped motors and gears
\item 4*4 keypad
\item Buzzer and power adaptor
\item Node MCU
\item LEDs and resistors
\item I2C LCD display
\end{itemize}
\subsection{Working}
Smart ID card is divided into 3 sub-systems\newline
1) Security System\newline
2) Payment System\newline
3) Attendance System\newline
\begin{itemize}
\item Security System \newline
The RFID reader communicates with the Arduino through the SPI protocol .The I2C
LCD communicates with the Arduino through the I2C protocol. The keypad is connected to Arduino. The 4X4 keypad has 8 connections but the last column of keypad is not required. We only require numbers for the password.\newline For powering the SIM900 module, 5V, 2A power adaptor is used. Once the SIM900 module is powered, the power light will light up and on pressing the power key, the status led lights up. Then the phone is paired with the module.\newline\newline

GSM Module:\newline
GSM is a mobile communication modem; it is stands for global system for mobile
communication (GSM). It is widely used mobile communication system in the world. GSM is an open and digital cellular technology used for transmitting mobile voice and data services.\newline
GSM module is used here since it can communicate with a mobile and the data which it receives can be processed and sent to the Arduino.\newline\newline
I2C Protocol:\newline
I2C is a serial protocol for two-wire interface to connect low-speed devices like microcontrollers, I/O interfaces and other similar peripherals in embedded systems.\newline
\begin{figure}[htbp]
\centering
\includegraphics[width=30pc]{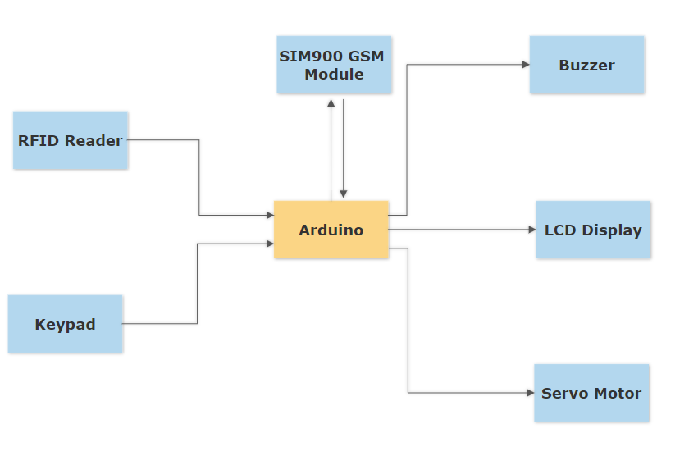}
\caption{Security System}
\label{fig_env1}
\end{figure}

\begin{figure}[htbp]
\centering
\includegraphics[width=30pc]{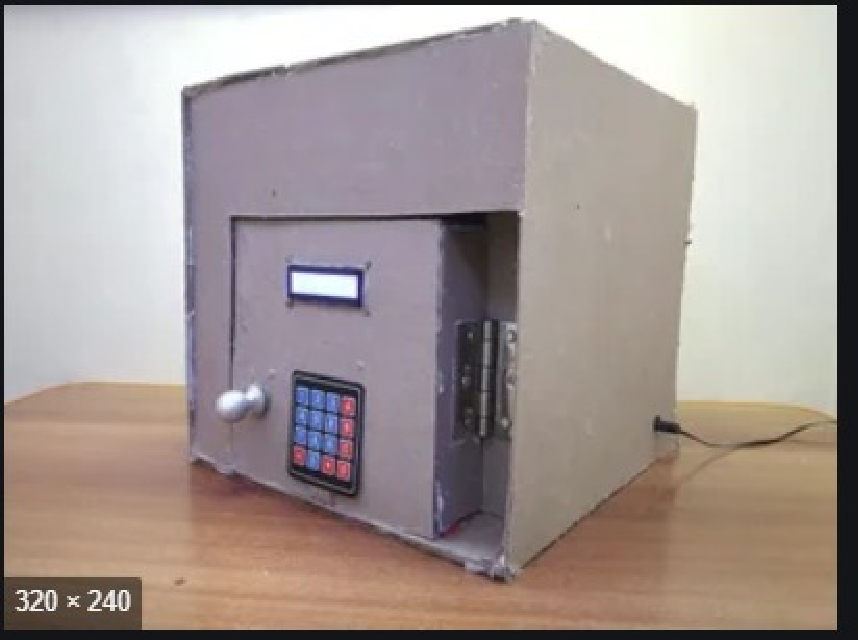}
\caption{Security System setup}
\label{fig_env1}
\end{figure}

\item Payment System:\newline 
The RFID reader communicates with the Node MCU through SPI protocol. The Node MCU is connected to a web server where the
data is stored. When the RFID card is scanned and the pin is entered , the balance amount is
displayed on the screen.\newline
Node MCU:\newline
This device is used instead of only Arduino UNO because Node MCU has a wi-fi module which can
be connected to the web server.
The ESP8266 can be controlled from local Wi-Fi network or from the internet (after port forwarding). The ESP-01 module has GPIO pins that can be programmed to control device/ execute a code through the internet. The module can be programmed using an Arduino through the serial pins (RX,TX).\newline

\begin{figure}[htbp]
\centering
\includegraphics[width=30pc]{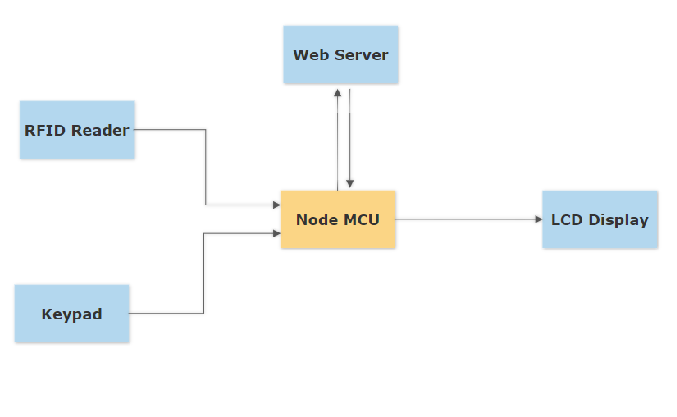}
\caption{Payment system}
\label{fig_env1}
\end{figure}
\begin{figure}[htbp]
\centering
\includegraphics[width=30pc]{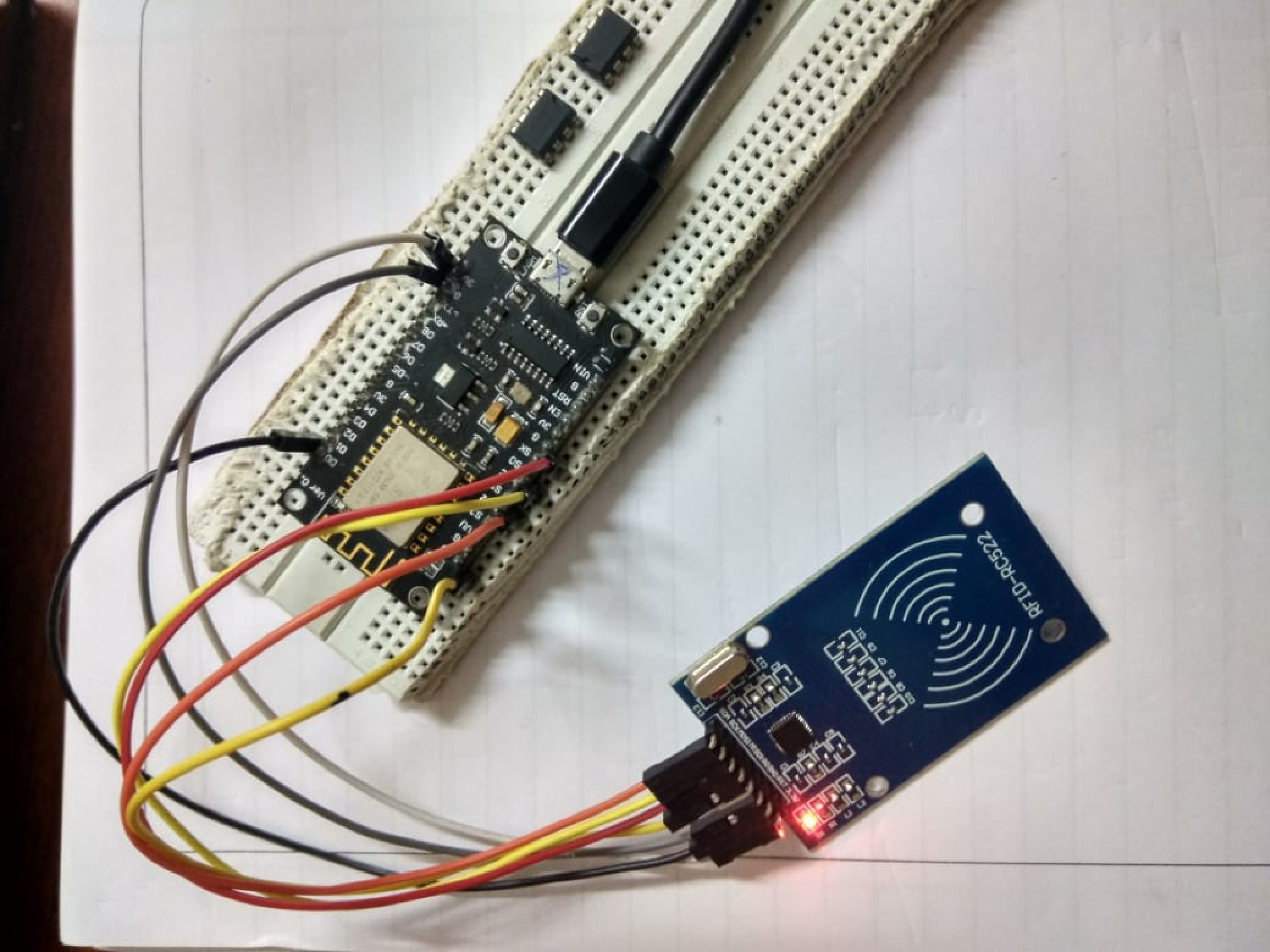}
\caption{Payment System setup}
\label{fig_env1}
\end{figure}

\begin{figure}[htbp]
\centering
\includegraphics[width=30pc]{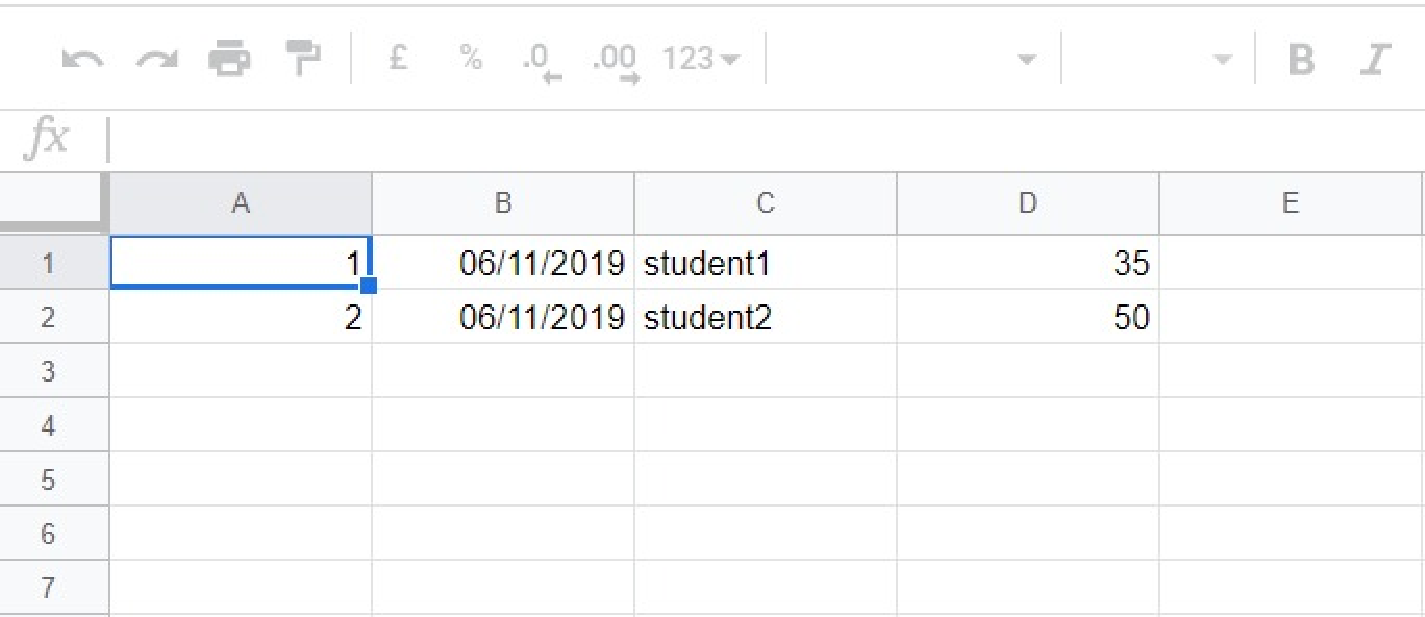}
\caption{Google docs for payment system}
\label{fig_env1}
\end{figure}
\item Attendance system:\newline
When the the ID is scanned on the RFID reader, the student name that is stored in the RFID card is printed on the serial monitor. It is made sure that the can't be registered twice by comparing it with already registered IDs. An external app is used to store the output from the serial monitor. The output can be saved on to the computer.\newline
\end{itemize}

 \begin{figure}[htbp]
\centering
\includegraphics[width=30pc]{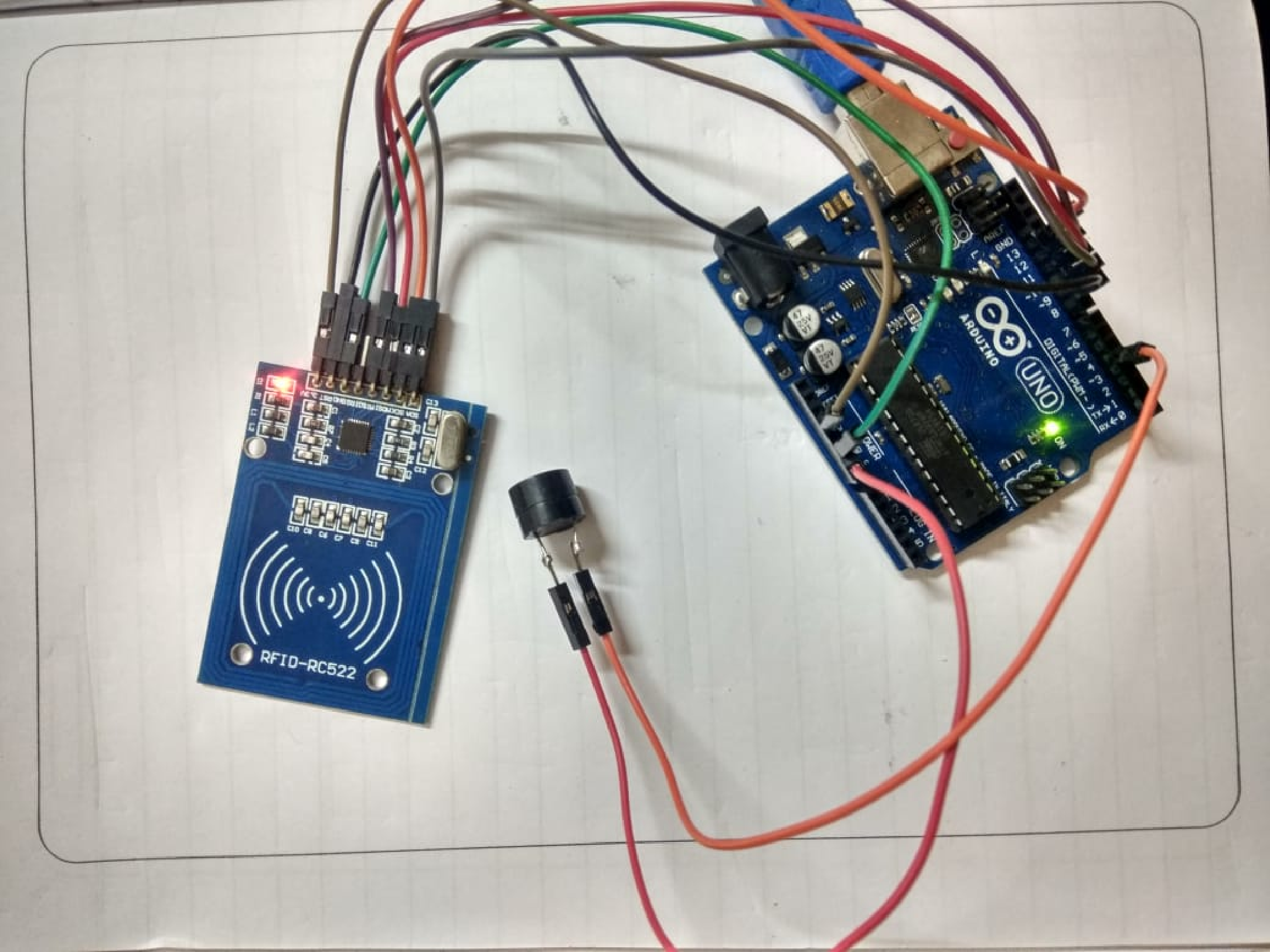}
\caption{Attendance System setup}
\label{fig_env1}
\end{figure}

\section{ {\textbf{Results and discussions}}}
\subsection{Security System}
The Door Lock security system was successfully implemented. When an authorized ID card is scanned onto the RFID reader and the correct password is entered onto the keypad, only then the door unlocks when the servo motor turns. Consequently a message is sent to the owner saying that the door is unlocked. After few seconds the door locks back, turning the servo motor to the original position \newline
When the owner is inside the room, he/she can use a switch which is present inside the room to unlock the door. Subsequently after few seconds the door locks backs, turning the servo motor back to the original position\newline
If in any case a wrong ID card or wrong password is entered. The whole system locks down and an alarm is buzzed using a buzzer. A message is sent to the owner saying that there was an attempt to breach the security system.\newline
The security system fails to detect an intruder when RFID card's ID is changed to the owners ID. It will also fail if the owner is negligent, revealing the password to others.
\subsection{Payment System}
when a ID is scanned in onto the RFID reader, the value that is stored in the RFID, is sent to the server via WIFI module through internet on to the data base with the date and time which is taken from the internet. This stored value can be changed by the vendor or the shopkeeper to the new balance amount. The changed balance amount is then updated in the ID card through the WIFI module ESP8266\newline
backdrop of this system is that the balance can be changed to a wrong value giving a wrong balance
\subsection{Attendance system}
The attendance system was successfully implemented. When an registered ID card is scanned onto the RFID reader, the ID card number is send to the database through the wifi module Node MCU. The data base saves the student's name, ID number on the database. This present list can be retrieved  from the database.\newline
As a fail safe for the above implemented method, the RFID reader reads the ID number of the card and compares it with the student register, if ID is present, it prints the student's name onto the serial monitor. An external app saves the logs of the serial monitor as text.\newline
This method would fail if some other student scans the card even if the owner is not present in the class. So the scanner must be monitored while the student is scanning on the RFID scanner

\section*{ {\textbf{Acknowledgment}}}

With immense pleasure we are presenting "Enhancing Room Security and Automating Class
Attendance Using ID Cards". As
a part of the curriculum of "Embedded Systems and Design" under the department of “Electronics and Communication Engineering, National Institute of Technology, Karnataka”. We wish to thank all people who gave us the unending support. We express my profound thanks to our Professor, Dr. Ramesh Kini M., And all those who have indirectly guided and helped us in the preparation of this project.  


\section*{}


\begin{thebibliography}{00}
\bibitem{b1} How RFID Works https://electronics.howstuffworks.com/gadgets/high-tech-gadgets/rfid.htm
\bibitem{b2} Specification of ESP8266
https://randomnerdtutorials.com/esp8266-adc-reading-analog-values-with-nodemcu/
\bibitem{b3} Data sheet ARDUINO UNO
https://www.farnell.com/datasheets/1682209.pdf

\end{thebibliography}
\end{document}